\documentclass[5p, times, switch]{elsarticle}

\usepackage{amsmath} 
\usepackage{amssymb} 
\usepackage{algorithm} 
\usepackage{algpseudocode} 
\usepackage[markup=colored,commandnameprefix=ifneeded]{changes}

\usepackage{amsmath}
\usepackage{amssymb}
\usepackage{amsfonts}
\usepackage{multirow}
\usepackage{longtable}
\usepackage{caption}
\usepackage{subcaption}
\usepackage{pifont}
\usepackage{array}
\usepackage{url}
\usepackage{xcolor}
\usepackage[colorlinks=true,allcolors=white]{hyperref}
\usepackage{makecell}
\usepackage{booktabs} 
\usepackage{stfloats}
\usepackage{svg} 

\usepackage{minted}

\newcolumntype{P}[1]{>{\centering\arraybackslash}p{#1}}

\usepackage{listings} 
\usepackage[most]{tcolorbox}

\begin{document}

\let\WriteBookmarks\relax
\def\floatpagepagefraction{1}
\def\textpagefraction{.001}





\title{PoC-Adapt: Semantic-Aware Automated Vulnerability Reproduction with LLM Multi-Agents and Reinforcement Learning-Driven Adaptive Policy}




%

\affiliation[1]{organization={Information Security Lab, University of Information Technology},
    city={Ho Chi Minh City},
    country={Vietnam}}
\affiliation[2]{organization={Vietnam National University},
    city={Ho Chi Minh City},
    country={Vietnam}}

\author[1,2]{Phan The Duy \corref{cor1}}
\ead{duypt@uit.edu.vn}

\author[1,2]{Khoa Ngo-Khanh}
\ead{khoank@uit.edu.vn}

\author[1,2]{Nguyen Huu Quyen}
\ead{quyennh@uit.edu.vn}

\author[1,2]{Van-Hau Pham}
\ead{haupv@uit.edu.vn}

\cortext[cor1]{Corresponding author}

\begin{abstract}

The rapid growth of software vulnerabilities and the increasing complexity of modern software ecosystems have made manual vulnerability reproduction and exploit validation increasingly impractical. While recent approaches leverage large language models (LLMs) and multi-agent pipelines to automatically generate proof-of-concept (PoC) exploits from vulnerability reports, existing systems often suffer from two fundamental limitations: unreliable validation based on surface-level execution signals and high operational cost caused by extensive trial-and-error during exploit generation. In this paper, we present PoC-Adapt, an end-to-end framework for automated PoC generation and verification, architected upon a foundation semantic runtime validation and adaptive policy learning. At the core of PoC-Adapt is a Semantic Oracle that validates exploits by comparing structured pre- and post-execution system states, enabling reliable distinction between true vulnerability exploitation and incidental behavioral changes. To reduce exploration cost, we further introduce an Adaptive Policy Learning mechanism that learns an exploitation policy over semantic states and actions, guiding the exploit agent toward effective strategies with fewer failed attempts. PoC-Adapt is implemented as a multi-agent system comprising specialized agents for root cause analysis, environment building, exploit generation, and semantic validation, coordinated through structured feedback loops. Experimenting on the CWE-Bench-Java and PrimeVul benchmarks shows that PoC-Adapt significantly improves verification reliability by 25\% and reduces exploit generation cost compared to prior LLM-based systems, highlighting the importance of semantic validation and learned action policies in automated vulnerability reproduction. Applied to the latest CVE corpus, PoC-Adapt confirmed 12 verified PoC out of 80 reproduce attempts at a cost of \$0.42 per generated exploit.

\end{abstract}

\begin{keyword}
Large Language Models \sep Vulnerability Reproduction \sep Proof-of-Concept \sep Reinforcement Learning  \sep Multi-Agent Systems \sep Exploit Generation \sep Agentic AI \sep Adaptive Policy
\end{keyword}



\maketitle

\section{Introduction} \label{sec_introduction}
The software industry is witnessing an unprecedented surge in vulnerabilities, driven by the complexity of modern supply chains, microservices architectures, and widespread adoption of open-source software (OSS). In 2025 alone, the number of disclosed Common Vulnerabilities and Exposures (CVEs) is projected to exceed 50,000, marking a 23\% increase from the previous year \cite{leverett2025q4forecast}. This rapid proliferation outpaces human capacity for manual assessment and remediation. Attackers exploit vulnerabilities within an average of 5 days, while organizations take approximately 38 days to patch them \cite{verizonDBIR2025}. Furthermore, supply chain attacks have risen by 30\%, often leveraging vulnerabilities as initial entry vectors \cite{verizonDBIR2025}. 

Compounding this issue, most vulnerability reports provide only concise descriptions without accompanying Proof-of-Concept (PoC) exploits, creating a reproducibility gap that hinders timely verification, patching, and risk assessment.
Meanwhile, PoCs are critical in the vulnerability management lifecycle, serving as verifiable demonstrations of exploitability. They enable developers to understand root causes, design effective patches, and integrate regression tests to prevent recurrence. For security teams, PoCs facilitate accurate risk prioritization based on real-world impact rather than abstract scores. However, manual PoC generation is labor-intensive, requiring deep expertise in code analysis, environment setup, and exploitation techniques, making it infeasible at scale \cite{brooks2018survey, vishnyakov2021survey, sapia2026scaling, zhao2025systematic}. On the other hand, traditional approaches to automated exploit generation (AEG), such as symbolic execution and constraint solving \cite{brumley2008apeg, avgerinos2011aeg, alhuzali2016chainsaw}, excel in formal analysis but lack generality across diverse ecosystems, languages, and vulnerability types. They often fail with ambiguous descriptions or complex interactions. 

Recent advancements in Large Language Models (LLMs) offer a promising alternative, leveraging their ability to process natural language, code, and tools for contextual reasoning in cybersecurity tasks such as vulnerability analysis \cite{mei2024arvo, simsek2025pocgen, pu2026patch, andersson2025poco} and penetration testing \cite{2024PentestGPT, luong2025xoffense, deng2026makes, wang2025ptfusion}. Vulnerability analysis systems like FaultLine \cite{nitin2025faultline} employ LLM agents for structured reasoning, identifying taint paths and constraints to generate PoCs. CVE-Genie \cite{ullah2025fromcve} extends this with a multi-agent pipeline for end-to-end reproduction, incorporating developer-critic loops to mitigate hallucinations. Despite these advances, significant challenges persist. Input contexts are often inadequately processed, leading to information loss or noise. LLMs are prone to hallucinations and biased inferences. Verification oracles rely on superficial signals such as flag checks or crashes, missing subtle semantic impacts. Moreover, heuristic trial-and-error in large action spaces incurs high computational costs without long-term adaptation from past exploits.

To address these gaps, we propose PoC-Adapt, an end-to-end framework for automated PoC synthesis and verification orchestrating multi-agent LLMs through adaptive policy learning. The framework is designed to improve both the reliability and efficiency of automated vulnerability reproduction. 
In particular, PoC-Adapt decomposes the overall task into a sequence of coordinated stages, including context retrieval, root cause analysis (RCA), environment setup, exploit generation, and semantic verification. 
In addition, a key innovation is the adaptive policy learning mechanism, modeling the exploitation process as a Markov Decision Process (MDP) and training a Double Deep Q-Network (DDQN) agent on exploitation logs to optimize action selection, minimizing heuristics and steps to success.

Our main contributions are as follows:

\begin{itemize}
    \item We introduce a Semantic Oracle for robust exploit validation based on structured state-differential analysis. By comparing pre- and post-execution system states, the oracle can reliably distinguish true vulnerability exploitation from incidental behavioral changes, overcoming the weaknesses of conventional crash-based or flag-based validation methods.
    
    \item We propose an Adaptive Policy Learning mechanism that formulates exploit generation as a MDP and trains a DDQN offline using exploitation logs. The learned policy guides the exploit agent toward more effective action sequences, reducing failed attempts, exploration overhead, and operational cost compared to heuristic-driven strategies.
    
    \item We design a tightly coordinated multi-agent orchestration pipeline with specialized agents for root cause analysis, environment setup, exploit generation, and semantic validation, enhanced by structured inter-agent feedback loops and context filtering. This modular architecture minimizes error propagation, improves traceability, and supports end-to-end automation across diverse vulnerability types.
\end{itemize}

The paper is structured as follows: Section~\ref{sec:background_related} provides theoretical foundations and reviews related work; Section~\ref{sec:method} details PoC-Adapt's architecture; Section~\ref{sec:implementation_experiments} presents implementation, experimental settings; whereas Section \ref{sec:results} gives experimental evaluation and result analysis. Next, the discussion about limitations and threats affecting to findings are mentioned in Section \ref{sec:threats_to_validity}. Finally, Section~\ref{sec:conclusion} concludes with limitations and future directions.

\section{Background and Related Work}
\label{sec:background_related}

\subsection{Background}

The rapid growth of software vulnerabilities in modern ecosystems, driven by microservices, complex supply chains, and extensive open-source dependencies, has created an urgent need for automated tools capable of synthesizing and verifying Proof-of-Concept (PoC) exploits. Large Language Model (LLM) agents have emerged as powerful tools for this task, extending beyond pure text generation through \emph{tool-use} capabilities that enable interaction with real environments \cite{schick2023toolformer}.

A foundational mechanism is \emph{Reasoning and Acting} (ReAct) \cite{yao2023react}, which structures LLM behavior into an observable cycle: \emph{thought} (natural-language reasoning), \emph{action} (tool invocation), and \emph{observation} (execution feedback). This loop significantly reduces hallucinations by grounding inferences in real observations, provides traceable reasoning traces, and bridges linguistic planning with concrete environmental manipulation, critical for reliable PoC generation where fabricated code paths or states are common failure modes.

To further improve reliability, self-verification techniques have been introduced. Self-Refine \cite{madaan2023selfrefine} enables iterative linguistic feedback and refinement of outputs. Reflexion \cite{shinn2023reflexion} maintains long-term verbal memory to adjust reasoning heuristics across trials. In multi-agent architectures, such as CVE-Genie \cite{ullah2025fromcve}, dedicated Critic agents provide cross-verification, categorizing feedback into output-level (self-generated) and interaction-based (environment or inter-agent) forms.

Multi-agent collaboration decomposes complex, long-horizon tasks into specialized roles with shared context and cross-checking, as demonstrated in PentestGPT \cite{deng2023pentestgpt} and VulnAgent \cite{wang2025vulagent}. While modular and robust against cascading errors, these systems incur higher latency and coordination overhead.

Verification oracles determine PoC validity beyond mere execution. Semantic oracles, inspired by smart contract verification \cite{chen2025smartpoc}, compare pre- and post-execution system states to confirm exploitation impact, offering greater fidelity than crash-based or flag-check oracles.

Reinforcement Learning (RL) formalizes sequential decision-making as a Markov Decision Process (MDP) $\mathcal{M} = \langle \mathcal{S}, \mathcal{A}, P, R, \gamma \rangle$, with the objective of maximizing expected discounted reward $J(\pi) = \mathbb{E}_{\pi} \Big[ \sum_{t=0}^{\infty} \gamma^t r_t \Big]$ \cite{sutton2018rl}. Our work employs offline, model-free, off-policy RL \cite{levine2020offline} to learn from static exploitation logs, avoiding expensive real-time interaction. For discrete action spaces (tool selection), Double Deep Q-Networks (DDQN) \cite{vanhasselt2016double} approximate action-value functions $Q(s,a)$ while mitigating overestimation bias.

\subsection{Related Work}

Automated exploit generation (AEG) has progressed from symbolic and constraint-based methods to LLM-driven approaches, each addressing aspects of the reproducibility gap where most disclosed vulnerabilities lack verifiable PoCs.


\paragraph{Traditional AEG} Early systems primarily relied on formal techniques such as symbolic execution, constraint solving, and binary diffing \cite{brumley2008apeg, avgerinos2011aeg}. Tools like APEG, AEG, and Chainsaw paved the way for automated reasoning in memory and web vulnerabilities \cite{brumley2008apeg, alhuzali2016chainsaw}. SemFuzz \cite{you2017semfuzz} applies semantics-guided fuzzing with runtime sanitizers. ARVO \cite{mei2024arvo} provides a large reproducible dataset of memory bugs in C/C++ OSS projects but remains passive in exploit synthesis. Recent studies have extended these concepts to handle complex exploit contexts. For instance, DEPA utilizes fuzzing and concolic execution to discover heap exploitation primitives \cite{liu2022depa}, while FLOWSTITCH leverages data-flow stitching to synthesize data-oriented attacks without hijacking the control flow \cite{hu2015automatic}. ARCANIST automatically constructs robust ROP and code-reuse gadget chains to bypass modern mitigations \cite{bailluet2025nothing}, and FIXX employs Code Property Graphs (CPG) to discover similar exploitable paths based on known examples \cite{thimmaiah2025fixx}. SAEG proposes an extensible state machine based on Exploit Graphs to handle multi-step exploitation procedures \cite{wu2024saeg}. Despite offering strong formal guarantees, these approaches often suffer from limited generalizability, poor handling of ambiguous natural language descriptions, and a lack of environment automation \cite{brumley2008apeg}. Furthermore, their verification oracles typically rely on implicit assertions like program crashes, which do not always guarantee successful exploitation \cite{shen2024revealing,bui2025systematic}.

\begin{table*}[t]
\centering
\caption{Comparison of related PoC Generation Frameworks.}
\label{tab:rw_framework_compact}
\resizebox{\textwidth}{!}{
\begin{tabular}{lcccccc}
\toprule
\textbf{Method} & \textbf{Approach} & \textbf{Oracle} & \textbf{Self-Critique} & \textbf{Adaptivity} & \textbf{Rebuild} & \textbf{Generalizability} \\
\midrule
APEG \cite{brumley2008apeg} & Symbolic Exec., Binary diffing & Off-the-shelf (BitBlaze) & N/A & N/A & No & No \\
AEG \cite{avgerinos2011aeg} & Symbolic Exec., Constraint solving & Execution (shell check) & N/A & N/A & No & No \\
Chainsaw \cite{alhuzali2016chainsaw} & Symbolic Exec., Constraint solving & Reachability & N/A & N/A & No & No \\
SemFuzz \cite{you2017semfuzz} & Semantic-guided fuzzing & AddressSanitizer & N/A & N/A & No & No \\
ARVO \cite{mei2024arvo} & Vulnerability reproduction & AddressSanitizer & N/A & N/A & Yes & No \\
PoCGen \cite{simsek2025pocgen} & LLM + static-dynamic & Specific validation & Validation-based & Heuristic & Yes & No \\
FaultLine \cite{nitin2025faultline} & LLM-based hierarchical reasoning & Runtime behavior (crash/state) & Feedback-driven loop & Heuristic & Yes & Yes \\
CVE-Genie \cite{ullah2025fromcve} & LLM multi-agent pipeline & Flag-check (CTF-style) & Developer-Critic loop & Heuristic & Yes & Yes \\
FLOWSTITCH \cite{hu2015automatic} & Data-flow stitching & Implicit assertions & N/A & N/A & No & No \\
ARCANIST \cite{bailluet2025nothing} & SMT-based code-reuse & Implicit assertions & N/A & N/A & No & No \\
Chit-chat \cite{caturano2025chit} & Dual-LLM + GDB & Execution (shell) & FSM matching & FSM Heuristic & Yes & Yes \\
PTFusion \cite{wang2025ptfusion} & Multi-agent + DKG & Raw Tool outputs & CoT logic & DKG Reasoning & No & Yes \\
DeepAttacker \cite{qu2025deepattacker} & Multi-agent + RAG & CALDERA/Exec & N/A & RAG Heuristic & No & Yes \\
\textbf{PoC-Adapt (Ours)} & \textbf{LLM multi-agent + RL} & \textbf{Semantic state-diff + runtime} & \textbf{Inter-agent + dev-critic} & \textbf{DDQN policy learning} & \textbf{Yes} & \textbf{Yes} \\
\bottomrule
\end{tabular}
}
\end{table*}

\paragraph{LLM and Multi-Agent based AEG} Recent works leverage Large Language Models (LLMs) to overcome the limitations of traditional formal methods, utilizing their capacity for contextual reasoning and code generation. Systems like FaultLine structure LLM reasoning into hierarchical data-flow analysis followed by feedback-driven PoC refinement \cite{nitin2025faultline}. PoCGen \cite{simsek2025pocgen} integrates LLMs with static taint analysis for NPM ecosystems, employing generate–validate–refine loops. To handle more complex tasks, the focus has shifted heavily toward Multi-Agent Systems (MAS). Chit-chat introduces a dual-LLM conversation (Llama 2 and ChatGPT) guided by finite state machines and GDB debuggers for buffer overflow exploitation \cite{caturano2025chit}. PTFusion employs a context-aware Dynamic Knowledge Graph (DKG) and the Model Context Protocol (MCP) to distribute tasks among specialized agents for web penetration testing \cite{wang2025ptfusion}. Similarly, DeepAttacker utilizes Retrieval-Augmented Generation (RAG) and the MITRE ATT\&CK framework within a multi-agent setup for breach and attack simulation \cite{qu2025deepattacker}. CVE-Genie implements a full end-to-end multi-agent pipeline with CTF-style verification and developer-critic loops \cite{ullah2025fromcve}.

Despite these advancements, existing LLM-based frameworks still face two fundamental limitations: unreliable validation based on surface-level execution signals and high operational costs caused by extensive trial-and-error exploration.

Table~\ref{tab:rw_framework_compact} compares these frameworks across core dimensions. As illustrated, PoC-Adapt distinguishes itself as the sole architecture uniting a semantic state-differencing oracle with DDQN-driven policy learning. While current multi-agent systems rely on static heuristics or RAG for adaptivity, PoC-Adapt dynamically adapts to varying exploitation contexts with high generalizability while providing robust validation.



\section{Proposed Method}
\label{sec:method}

\begin{figure*}[!h]
    \centering
    \includegraphics[width=1.75\columnwidth]{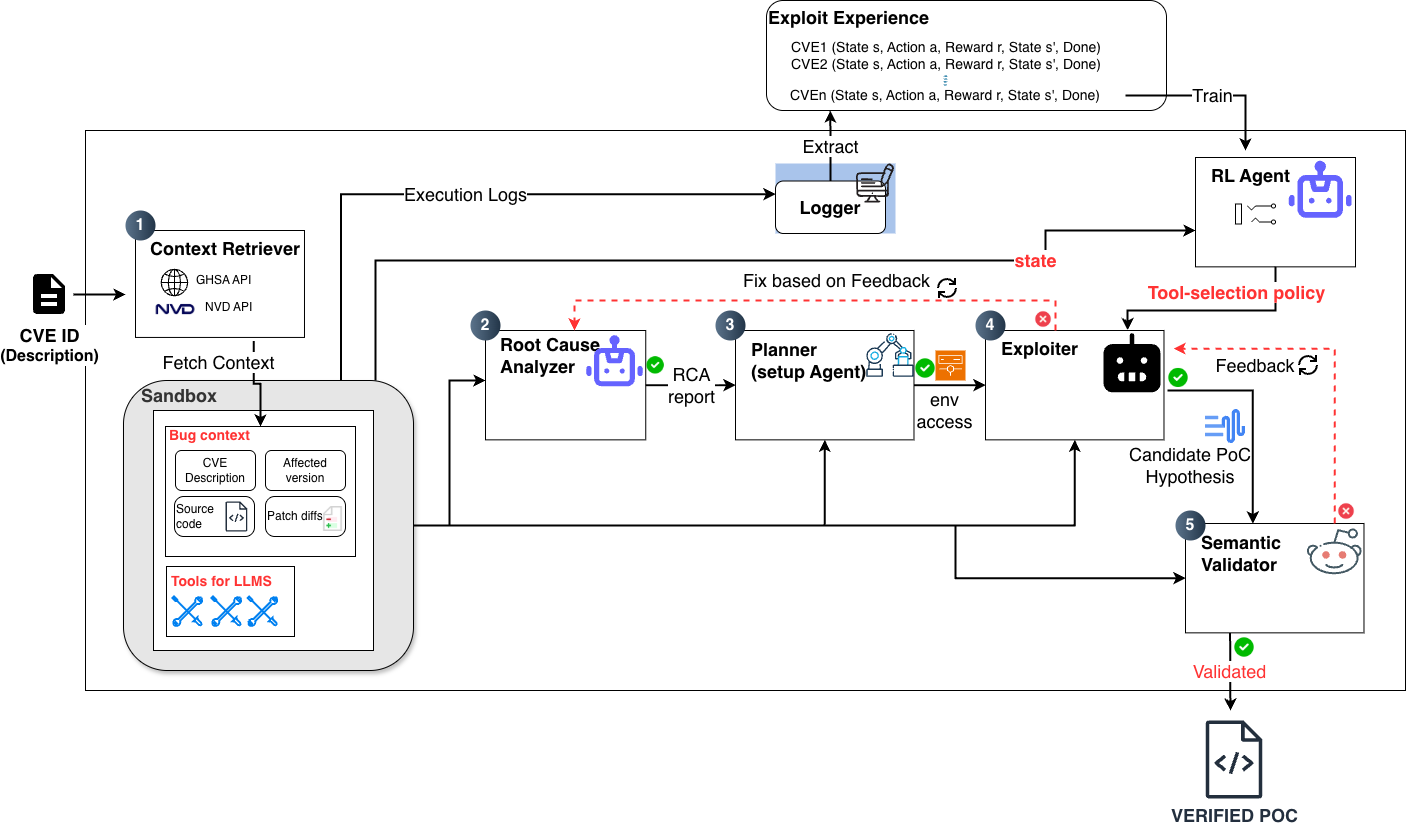}
    \caption{Overall architecture of PoC-Adapt and data processing flow.}
    \label{fig:pipeline_architecture}
\end{figure*}

To address the limitations of prior automated PoC generation systems, such as FaultLine \cite{nitin2025faultline} and CVE-Genie \cite{ullah2025fromcve}, particularly unreliable runtime verification and high failure rates caused by heuristic trial-and-error in large action spaces, we introduce PoC-Adapt, an end-to-end framework for automated PoC synthesis and verification. Unlike existing approaches that often rely on coarse execution signals to determine exploit success, PoC-Adapt is designed to reason over the semantic effects of an exploit attempt on the target system. Its design is grounded in two complementary components. The first is a Semantic Oracle, which validates exploit attempts by analyzing structured differences between pre- and post-execution system states, thereby enabling more reliable discrimination between true exploitation outcomes and unrelated runtime side effects. The second is an Adaptive Policy Learning mechanism, which leverages feedback from prior interactions to improve action selection during exploitation, reducing ineffective exploration and lowering overall generation cost. The complete framework, shown in Fig.~\ref{fig:pipeline_architecture}, organizes the PoC generation workflow into five sequential stages handled by specialized agents, with structured feedback loops connecting these stages to preserve traceability and continuously refine downstream decisions.

\subsection{The Overview of System Architecture}
PoC-Adapt processes inputs comprising a CVE-ID or GHSA-ID and a natural-language vulnerability description. The pipeline generates a verified PoC or flags failure with diagnostic states.

The pipeline is designed as a sequential workflow to ensure logical progression from raw vulnerability data to verified exploitation, minimizing error propagation and enabling efficient debugging. This staged approach draws from traditional vulnerability analysis pipelines but incorporates agent-based modularity for flexibility. The five stages are:

\begin{itemize}
    \item \textbf{Context Retrieval} extracts CVE details (description, source code, patch diffs, affected versions) from NVD/GHSA APIs, forming a \emph{bug context} $C = \langle desc, repo, patch, \\affected\_ver \rangle$. This stage is crucial as it grounds all subsequent reasoning in verifiable data, reducing hallucinations by limiting assumptions.
    \item \textbf{Root Cause Analysis} localizes the bug, identifies taint paths, and extracts constraints, producing $R = \langle loc, sink, \\entry, paths, constraints, steps \rangle$. By focusing on data flow and control constraints, this stage provides a structured foundation for exploitation, addressing the ambiguity in vague reports.
    \item \textbf{Environment Setup} configures a reproducible vulnerable environment, yielding $P = \langle envSpec, buildCmds, run\\Cmds, accessInfo \rangle$. Isolation via Docker ensures safety and reproducibility, essential for testing without risking host systems.
    \item \textbf{Exploit Generation} synthesizes candidate PoC $E$ and hypothesis $H$ for expected impact. The hypothesis formalizes verifiable outcomes, bridging generation and validation.
    \item \textbf{Semantic Verification} validates $E$ against $H$, returning $V = \langle verdict, PoC \rangle$. This final gatekeeper prevents false positives through semantic checks.
\end{itemize}

Failure at any stage such as exceeding budget results in diagnostic labels like NOT\_VALIDATED, allowing targeted debugging.

Furthermore, PoC-Adapt also employs a tightly coordinated pipeline of four role-specialized agents with controlled tool access for modularity and safety. Instead of operating independently, the output of one agent directly becomes the context and prerequisite for the next, forming a closed-loop process from theoretical analysis to practical validation. Role specialization follows the principle of separation of concerns, assigning each agent a focused task to reduce cognitive load on LLMs and minimize errors from multitasking.

\begin{itemize}
    \item \textbf{Root Cause Analyzer Agent:} The pipeline begins with the RCA agent. Receiving the bug context $C$ as input, this agent plays a core role in bug localization and taint analysis. It is designed to mimic expert vulnerability research by systematically tracing data flows, using tools to avoid manual code navigation errors. It backtraces from the bug location (sink) to the entry points and extracts control flow constraints. The output is the RCA report $R$, which acts as a "theoretical map" providing detailed information on how input data can bypass checks to reach the vulnerable code snippet. Pseudocode is provided in Algorithm~\ref{alg:rca_agent}.

    \begin{algorithm}[t]
    \caption{RCA Agent: Vulnerability Analysis}
    \label{alg:rca_agent}
    \begin{algorithmic}[1]
    \Statex \textbf{Input:} Bug context $C$, feedback $fb$ (optional)
    \Statex \textbf{Output:} RCA report $R$
    \State $S \gets \textsc{CollectSignals}(C)$
    \State $cand \gets \textsc{CodeSearch}(S)$
    \State $loc \gets \textsc{LocalizeBug}(cand, C.patch)$
    \State $sink \gets \textsc{IdentifySink}(loc)$
    \State $entry \gets \textsc{FindEntryPoints}(sink)$
    \State $paths \gets \textsc{TraceTaintPaths}(entry, sink)$
    \State $constraints \gets \textsc{ExtractGuards}(paths)$
    \If{$fb \neq \emptyset$}
        \State $(loc,paths,constraints) \gets \textsc{RefineWithFeedback}(loc,paths,constraints,fb)$
    \EndIf
    \State $steps \gets \textsc{SummarizeTriggerSteps}(entry,paths,constraints)$
    \State \Return $R=\langle loc,sink,entry,paths,constraints,steps\rangle$
    \end{algorithmic}
    \end{algorithm}

    \item \textbf{Planner Agent:}Once the theoretical map $R$ is established, the system requires a practical environment for verification. The Planner agent takes $R$ and $C$ to automatically set up the vulnerable environment. It iteratively refines setups based on execution feedback, designed to handle diverse ecosystems by analyzing build systems and documentation to synthesize setup plans. The output is the Planner report $P$, which ensures that the environment is not only successfully built but also exposes accessible endpoints or ports for the next agent to interact with. 


    \item \textbf{Exploiter Agent:} Armed with the theoretical insights from $R$ and the practical environment from $P$, the Exploiter agent proceeds with exploit development. This agent leverages its toolset to interact with the environment $P$ and satisfy the constraints identified in $R$ to craft targeted exploits. Upon success, it generates not only the expected exploit script $E$ but also an impact hypothesis $H$. This hypothesis clearly describes the expected state changes in the system if the exploit is successful, serving as a mandatory prerequisite for the semantic validation phase.

    \item \textbf{Validator Agent:} In the final step, to avoid misleading evaluations based solely on return codes (as discussed in Section~\ref{sec_introduction}), the Validator acts as an independent verification unit. Taking $E$ and $H$ from the Exploiter as input, this agent directly executes the exploit against the environment. It observes and verifies the actual system state changes against the hypothesis $H$. The final output $V=\langle \texttt{VALIDATED}/\texttt{NOT\_VALIDATED}, \texttt{PoC}\rangle$ formally confirms the validity of the exploit. Details are in Section~\ref{sec:semantic_oracle}.

    \item \textbf{Agent Coordination and Feedback Loops:}
    To ensure the pipeline operates as a tightly coordinated, closed-loop system, PoC-Adapt leverages self-verification and structured inter-agent feedback mechanisms. These mechanisms collectively minimize hallucinations, propagate semantic errors backward for refinement, and significantly reduce exploration cost without any human-in-the-loop intervention.
    
    Self-verification, inspired by Reflexion~\cite{shinn2023reflexion}, enables each agent (particularly the Exploiter) to internally critique and iteratively refine its own outputs—such as PoC candidates and associated hypotheses—against the constraints and taint paths identified in the RCA report. This intra-agent reflection loop bounds the number of refinement iterations and grounds reasoning in prior execution feedback.
    
    Complementing this, inter-agent feedback channels—most notably between the Exploiter and the Validator (Semantic Oracle), as well as back-propagation to the RCA agent allow downstream semantic verification signals to drive upstream refinement. For instance, when the Semantic Oracle detects a mismatch in pre- and post-execution system states ($\Delta \neq \emptyset$), it returns structured, contextual feedback (including failure category, affected state attributes, and suggested adjustments) to the Exploiter. This targeted feedback prevents error propagation across pipeline stages and eliminates the need for redundant per-agent critic modules, as seen in some prior multi-agent designs~\cite{talebirad2023multiagent}.
    
    Crucially, both self-verification outcomes and inter-agent feedback signals are logged as trajectories and incorporated into the offline replay buffer, directly fueling the DDQN-based adaptive policy learning module (Section~\ref{sec:adaptive_learning}). This integration closes the loop between reliable semantic validation and efficient long-horizon exploration, constituting a key differentiator from prior heuristic-driven LLM-based exploit generation frameworks.
    
    \begin{figure}[t]
        \centering
        \includegraphics[width=\columnwidth]{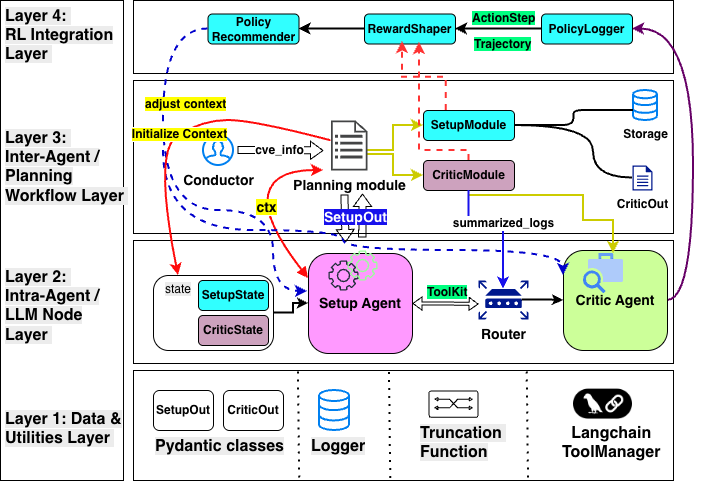} 
        \caption{Self-verification mechanism.}
        \label{fig:self_verify}
    \end{figure}

    \item \textbf{Tool Design and Controlled Allocation:}
    To effectively execute this orchestrated workflow, agents must interact with external systems. However, to guarantee safety and prevent arbitrary operations, such as code injection or resource exhaustion, tools are designed as a controlled, sandboxed set. Rather than granting global access, tool allocation follows the principle of least privilege, directly mapping to each agent's specific role in the pipeline. Tool definitions for agents are provided in Table~\ref{tab:tools_definition}.
    0 
    \begin{table}[t]
    \centering
    \caption{Tool Definitions}
    \label{tab:tools_definition}
    \resizebox{\columnwidth}{!}{
    \begin{tabular}{lp{6cm}}
    \toprule
    \textbf{Tool} & \textbf{Function} \\
    \midrule
    \texttt{get\_file} & Read file content in workspace \\
    \texttt{write\_to\_file} & Write/create new file \\
    \texttt{execute\_ls\_command} & List project files/directories \\
    \texttt{execute\_linux\_command} & Execute Linux commands \\
    \texttt{find} & Search files in project \\
    \texttt{grep} & Search file content via regex \\
    \texttt{semantic\_code\_search} & Semantic similarity code search \\
    \texttt{setup\_environment} & Setup application environment \\
    \texttt{rebuild\_env} & Rebuild on errors \\
    \texttt{run\_poc} & Execute PoC and capture output \\
    \texttt{refine\_poc} & Refine PoC via feedback \\
    \texttt{dynamic\_trace} & Trace execution for debugging \\
    \texttt{test\_exploit\_condition} & Test vulnerability triggers \\
    \texttt{inspect\_runtime\_state} & Inspect runtime state \\
    \texttt{analyze\_error\_output} & Analyze errors/outputs \\
    \texttt{set\_environment\_variable} & Set environment variables \\
    \bottomrule
    \end{tabular}
    }
    \end{table}
\end{itemize}

\subsection{Semantic Oracle}
\label{sec:semantic_oracle}
As discussed in Section \ref{sec_introduction}, the most critical challenge in automated exploit generation is accurately determining the validity of the generated PoC. Relying solely on superficial indicators—such as program exit codes, network response statuses, or raw console outputs is highly prone to false positives, often resulting from LLM hallucinations where the model incorrectly assumes success. 
To address this limitation and ensure absolute system reliability, we propose a Semantic State Differencing Oracle. This mechanism completely shifts the verification paradigm from analyzing the exploit's output to analyzing the target system's state. It forces the Validator agent to observe, measure, and compare the actual system states before and after the exploit execution.

\subsubsection{Verification Workflow}
The Semantic Oracle operates through a rigorous three-phase pipeline, heavily utilizing the impact hypothesis $H$ generated by the Exploiter agent:

\begin{itemize}
    \item Phase 1: Pre-Execution State Profiling (Pre-Check). Before the PoC is executed, the Validator agent parses the hypothesis $H$ to understand the expected impact of the vulnerability. Based on this, it actively probes the isolated environment to capture a baseline state snapshot ($S_{\text{pre}}$). This snapshot may target specific directories, hash values of sensitive files, environment variables, or database records that the exploit claims to alter.
    
    \item Phase 2: Isolated Execution (Execute PoC). The candidate exploit $E$ is executed within the sandboxed vulnerable environment. During this phase, all execution logs, error traces, and exit statuses are comprehensively recorded. If the exploit crashes or fails at the syntax level, the execution logs are immediately formatted as feedback for refinement.
    
    \item Phase 3: Post-Execution Profiling and Semantic Differencing (Post-Check). Upon completion of the execution, the Validator recaptures the exact metrics defined in Phase 1 to obtain the post-execution state ($S_{\text{post}}$). The oracle then computes the semantic difference between the two states:
    \begin{equation}
    \Delta = \text{Analyze}(S_{\text{post}}, S_{\text{pre}})
    \end{equation}
\end{itemize}

The Validator agent analyzes the resulting $\Delta$ to identify concrete system anomalies, such as overwritten configurations, data exfiltration artifacts, or unauthorized privilege escalations. The PoC is only considered valid if the observed state changes $\Delta$ strictly match the theoretical impact described in the hypothesis $H$. By grounding the verification in observable, post-exploitation system realities, the Semantic Oracle effectively eliminates LLM hallucinations and guarantees the reliability of the generated exploits.

Algorithm~\ref{alg:semantic_oracle_impl} provides a comprehensive breakdown of how the Semantic Oracle operates. The system features a constrained refinement loop, capped at a maximum of $B$ iterations, which operates alongside the Exploiter agent. If an exploit is unsuccessful during execution or fails the semantic state verification, the Oracle produces highly specific, context-driven feedback. This feedback, denoted as $fb$, explains the exact reasons behind the PoC failure to the Exploiter, thereby facilitating continuous enhancement.

\begin{algorithm}[t]
\caption{Semantic Oracle Verification}
\label{alg:semantic_oracle_impl}
\begin{algorithmic}[1]
\Statex \textbf{Input:} Exploit $E$, hypothesis $H$, refinement budget $B$
\Statex \textbf{Output:} Verdict $V$, PoC if validated
\State $k \gets 0$; $fb \gets \emptyset$
\While{$k < B$}
    \State $\Pi \gets \textsc{BuildPrompt}(E, H, fb)$
    \State $C \gets \textsc{LLM\_GENERATE}(\Pi)$
    \State $S_{\text{pre}} \gets \textsc{PreCheck}(C)$
    \State $R \gets \textsc{ExecutePoC}(E)$
    \If{$R.status \neq \texttt{OK}$}
        \State $fb \gets \textsc{MakeFeedback}(\texttt{"EXECUTE\_FAIL"}, R)$
        \State $k \gets k + 1$
        \State $(E,H) \gets \textsc{RequestRefineFromExploiter}(fb)$
        \State \textbf{continue}
    \EndIf
    \State $S_{\text{post}} \gets \textsc{PostCheck}(C)$
    \State $\Delta \gets \textsc{AnalyzeDelta}(S_{\text{post}}, S_{\text{pre}})$
    \If{$\textsc{Match}(\Delta, H)$}
        \State \Return $(\texttt{VALIDATED}, E)$
    \Else
        \State $fb \gets \textsc{MakeFeedback}(\texttt{"NOT\_MATCH"}, \Delta)$
        \State $k \gets k + 1$
        \State $(E,H) \gets \textsc{RequestRefineFromExploiter}(fb)$
    \EndIf
\EndWhile
\State \Return $(\texttt{NOT\_VALIDATED})$
\end{algorithmic}
\end{algorithm}

\subsubsection{State Profiling}
\label{subsec:state-profiling}

The profiling process operates in three conceptual phases integrated directly into Algorithm~\ref{alg:semantic_oracle_impl}: (i) pre-execution baseline capture, (ii) isolated PoC execution, and (iii) post-execution analysis. These states are then compared to determine the semantic difference \(\Delta\), which serves as the definitive criterion for determining exploit validity.

Instead of relying on rigid heuristics or hardcoded mathematical set differences, the formal semantic differencing is delegated to a Large Language Model (LLM) acting as an autonomous judge. Given pre- and post-execution states \(S_{\text{pre}}\) and \(S_{\text{post}}\), the semantic difference \(\Delta\) is synthesized as a detailed reasoning narrative. The oracle is strictly prompted to compare these states and explicitly explain how the observed changes relate to the hypothesis.

The matching condition against the exploit hypothesis \(H\) (line~15 of Algorithm~\ref{alg:semantic_oracle_impl}) is then formally evaluated to yield a definitive boolean verdict:
\begin{equation}
    \text{Match}(\Delta, H) \triangleq \text{LLM\_judge}(\Delta, H) \in \{\text{True}, \text{False}\}
\end{equation}
where the evaluation strictly demands that the observed semantic changes logically prove the theoretical hypothesis \(H\).

In summary, State Profiling elevates the system from heuristic trial-and-error to a rigorously semantic and adaptive exploitation framework. For the Semantic Oracle, it fundamentally shifts the verification paradigm from unreliable surface-level signals (such as crashes or console outputs) to observable system impacts evaluated through structured LLM reasoning. This effectively minimizes hallucination-induced false positives and guarantees the semantic correctness of every verified PoC. Furthermore, the explicit reasoning narratives and validation verdicts supply rich trajectory data that enrich the state representations for the Adaptive Policy Learning module. This data enables the underlying Double DQN (DDQN) recommender to learn more effective policies, tightening the inter-agent refinement loop with precise, actionable feedback.

\subsection{Adaptive Policy Learning}
\label{sec:adaptive_learning}
Motivated by the limitations identified in our earlier analysis, we observe that the major challenges when operating a multi-agent pipeline based purely on LLMs are tool invocation costs, instability, and most notably, the severe risk of token explosion. Because LLMs are typically accessed via APIs and lack long-term memory to internalize lessons from past executions, the agents can easily fall into infinite trial-and-error loops. They often perform redundant exploratory actions or repeat the exact same mistakes when stuck. This not only degrades scalability but also causes token costs to skyrocket.

To mitigate this problem, we design an adaptive policy learning mechanism from execution logs using a RL agent. This mechanism acts as a strategic navigator, guiding the Exploiter Agent to make smarter and more optimal tool selections over time. The core concept is to shift away from letting the LLM dictate the action sequence entirely based on free-form reasoning, which easily leads to rambling and token explosion. Instead, the system fully leverages log data from historical exploitation trajectories—including both successful and failed attempts. By modeling this process, the RL-agent learns a policy $\pi(a \mid s)$ that explicitly prioritizes actions with a high probability of successful verification while aggressively pruning useless exploration branches. Consequently, the system can generate PoCs significantly faster, minimize redundant interactive steps, save computational resources, and thoroughly prevent token explosion.

\begin{figure}[h]
    \centering
    \includegraphics[width=\columnwidth]{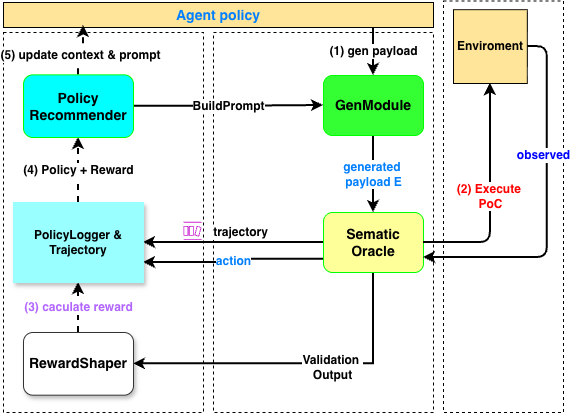} 
    \caption{Adaptive policy learning mechanism.}
    \label{fig:policy}
\end{figure}

Figure~\ref{fig:policy} details the deployment of our proposed Policy Learning Mechanism within the generation pipeline. At each step, rather than letting the LLM think unconstrained, the RL layer evaluates the current state and proposes the optimal macro-action/tool. The LLM is then restricted to executing that specific tool with appropriate parameters, effectively bounding the search space and ensuring discipline during code generation.

\subsubsection{MDP Formulation}
\label{sec:mdp}

In this framework, the decision-making process of the Exploiter Agent is formulated as a MDP consisting of:
\begin{itemize}
    \item State $s_t$: Summarizes the immediate context and the current progress of the exploitation.
    \item Action $a_t$: Represents the specific tool or macro-action selected by the agent.
    \item Reward $r_t$: Serves as the evaluative feedback provided by the environment.
    \item Next State $s_{t+1}$: Denotes the resulting system condition after the action is executed.
\end{itemize}
Furthermore, from the execution logs, we reconstruct trajectories $\tau=(s_0,a_0,s_1,\ldots,s_T)$ and train the policy using value-based RL to approximate the $Q(s,a)$ function. This allows the agent to "remember" costly lessons from past failures and deduce the optimal action trajectory.

\subsubsection{State Space Definition}
\label{sec:state_space}
\begin{table}[H]
\centering
\caption{State Space Features}
\label{tab:state_space}
\resizebox{\columnwidth}{!}{
\begin{tabular}{llcc}
\toprule
\textbf{Feature} & \textbf{Type} & \textbf{Domain} & \textbf{Meaning} \\
\midrule
phase & Categorical & $\{0,1,2,3,4\}$ & Pipeline stage \\
cwe\_type & Categorical & $\{0,\dots,K_{\text{cwe}}-1\}$ & CWE group index \\
tool\_diversity & Continuous & $[0,1]$ & Tool diversity ratio \\
error\_rate & Continuous & $[0,1]$ & Failure rate to date \\
iteration & Discrete & $\{0,\dots,T_{\max}\}$ & Current step \\
last\_tool & Categorical & $\{0,\dots,K_{\text{tool}}-1\}$ & Previous action \\
last\_success & Binary & $\{0,1\}$ & Previous success \\
error\_pattern & Categorical & $\{0,\dots,K_{\text{err}}-1\}$ & Recent error type \\
has\_poc\_written & Binary & $\{0,1\}$ & PoC generated flag \\
auth\_required & Binary & $\{0,1\}$ & Authentication needed \\
sandboxed & Binary & $\{0,1\}$ & Sandboxed environment \\
sink\_hit & Binary & $\{0,1\}$ & Sink triggered \\
partial\_success & Binary & $\{0,1\}$ & Partial progress \\
\bottomrule
\end{tabular}
}
\end{table}

Because the system aims to generalize across a wide variety of vulnerabilities, defining an exhaustive state space is challenging. 
We derived our state criteria from empirical testing aimed at minimizing redundant data. This approach effectively filters out noise for the RL agent by isolating indicators that are distinctly observable and frequently logged during execution.
The state $s_t$ is represented by 13 features extracted directly from the logs in Table~\ref{tab:state_space}. Categorical and binary features are encoded via vocabulary indices, while continuous features are normalized to $[0,1]$ to stabilize training.
The variables are defined such that $K_{\text{cwe}}$ is the number of CWE groups, $K_{\text{tool}}$ is the number of tools, and $K_{\text{err}}$ corresponds to the error patterns. The variable $T_{\max}$ establishes the maximum allowable steps, functioning as a hard boundary to strictly eliminate token explosion.

\subsubsection{Action Space Definition}
\label{sec:action_space}
\begin{table}[H]
\centering
\caption{Action Space Features}
\label{tab:action_space}
\resizebox{\columnwidth}{!}{
\begin{tabular}{clp{5cm}}
\toprule
\textbf{ID} & \textbf{Action} & \textbf{Function} \\
\midrule
0 & submit\_and\_verify & Verify PoC per CVE criteria \\
1 & execute\_command & Run system commands for reconnaissance \\
2 & read\_file & Read source/config files \\
3 & search\_code & Search code for vulnerability patterns \\
4 & setup\_environment & Configure target application \\
5 & analyze\_runtime & Dynamic runtime analysis \\
6 & write\_exploit & Generate initial PoC \\
7 & modify\_exploit & Refine existing PoC \\
8 & run\_exploit & Execute PoC and observe \\
\bottomrule
\end{tabular}
}
\end{table}

The action space $\mathcal{A}$ of the Exploiter Agent consists of core \textbf{tools (macro-actions)}. By confining the actions to 9 essential operations (Table~\ref{tab:action_space}), the state-action space is kept highly manageable. This constraint helps the RL algorithm converge faster and heavily mitigates hallucinated commands from the LLM.

\subsubsection{Reward Function}
\label{sec:reward}

The reward function $\mathcal{R}(s_t,a_t,s_{t+1})$ provides signals to guide the agent. To accelerate the learning process, the rewards are designed to be dense based on logs: the agent receives immediate feedback after every tool execution rather than waiting until the end of the entire exploitation attempt.

\begin{itemize}
    \item \textbf{Immediate Step Reward:} At each time step:
    \begin{equation}
    r(s_t, a_t, s_{t+1}) = 
    \begin{cases}
    r_{\text{success}}(a_t) & \text{if } s_{t+1}.\text{last\_success} = 1, \\
    r_{\text{failure}}(a_t) & \text{if } s_{t+1}.\text{last\_success} = 0.
    \end{cases}
    \end{equation}
    The weights $r_{\text{success}}$ and $r_{\text{failure}}$ are tuned to heavily penalize actions that cause repetitive errors, forcefully steering the agent toward more productive tools.

    \item \textbf{Terminal Reward:}
    An episode terminates immediately when a PoC is successfully verified or the $T_{\max}$ threshold is reached:
    \begin{equation}
    r_{\text{terminal}}(s_T) = 
    \begin{cases}
    +25 & \text{if the PoC is successfully verified}, \\
    -10 & \text{otherwise such as budget exhaustion}.
    \end{cases}
    \end{equation}

    \item \textbf{Trajectory Return:}
    For a trajectory $\tau = (s_0, a_0, s_1, ..., s_T)$:
    \begin{equation}
    R(\tau) = \sum_{t=0}^{T-1} r(s_t, a_t, s_{t+1}) + r_{\text{terminal}}(s_T).
    \end{equation}
    
    The agent's objective is to learn a policy $\pi$ that maximizes the expected return:
    \begin{equation}
    J(\pi) = 
    \mathbb{E}_{\tau \sim \pi}
    \left[
    \sum_{t=0}^{T-1} \gamma^t \, r(s_t, a_t, s_{t+1}) 
    + \gamma^T r_{\text{terminal}}(s_T)
    \right],
    \end{equation}
    {where $\gamma \in (0,1)$ is the discount factor. This heavily encourages the agent to achieve the verification goal as quickly as possible to secure unattenuated rewards, directly counteracting the token-wasting tendencies of lengthy, meandering LLM reasoning.}
\end{itemize}

\subsubsection{Policy Training Algorithm}

We employ the DDQN \cite{vanhasselt2016double} to learn the action-selection policy from the log repository. DDQN is highly suitable for optimizing our code generation pipeline for three main reasons:
 \begin{itemize}
     \item \textbf{Maximizing Existing Logs (Offline/Off-policy Learning):} Generating interactive data directly with the environment is prohibitively expensive (requiring container setups, dependency installations, etc.). Therefore, DDQN utilizes \emph{batch offline learning}: it reuses a replay buffer $\mathcal{D}$ containing experience tuples $(s_t,a_t,r_t,s_{t+1},done)$ extracted from historical logs. The agent learns to systematically avoid documented "dead-ends" that waste resources.

     \item \textbf{Stable and Fast Convergence (Reduced Overestimation Bias):} DDQN decouples action selection from action evaluation, effectively mitigating the overestimation of Q-values common in standard DQN. The TD target is computed as:
    \begin{equation}
    y_t = r_t + \gamma \, Q_{\theta^-}\!\Big(s_{t+1}, \arg\max_{a'} Q_{\theta}(s_{t+1}, a')\Big).
    \end{equation}
    This allows the agent to more accurately assess the underlying risk of each tool, make decisive choices, and swiftly escape execution bottlenecks.

    \item \textbf{Optimization Across Complex Feature Spaces:} The neural network in DDQN approximates the $Q(s,a)$ function over the 13-dimensional state space. This synthesizes a holistic, global view of the exploitation progress for decision-making, rather than relying exclusively on the finite and easily distracted text context of the LLM.
 \end{itemize}

\section{Implementation and Experiment Settings}
\label{sec:implementation_experiments}

This section presents a comprehensive evaluation of PoC-Adapt. We organize the experiments around six research questions (RQs), describe the datasets, implementation details, experimental setup, and evaluation metrics. Results are then reported and analyzed to rigorously assess effectiveness, practicality, generalizability, efficiency, adaptability, and robustness.

\subsection{Research Questions}

To systematically evaluate PoC-Adapt, we define the following research questions:

    
    
    
    
    

\begin{itemize}
    \item \textbf{RQ1 (Effectiveness)}: How does PoC-Adapt perform in automated vulnerability reproduction compared to state-of-the-art baselines on standardized benchmarks?
    
    \item \textbf{RQ2 (Practicality)}: What is the end-to-end reproduction success rate of PoC-Adapt in real-world software environments, and at which stages of the pipeline do critical failures predominantly occur?
    
    \item \textbf{RQ3 (Generalizability)}: How effectively can the system generalize its reproduction capabilities across diverse vulnerability types (CWE categories) and varying severity levels?
    
    \item \textbf{RQ4 (Efficiency)}: What is the computational overhead (in terms of token consumption and monetary cost) required by PoC-Adapt, and how are these costs distributed across its operational stages?
    
    \item \textbf{RQ5 (Adaptability \& Ablation)}: To what extent does the adaptive policy learning mechanism contribute to the overall success rate and efficiency compared to non-adaptive approaches?
    
    \item \textbf{RQ6 (Robustness)}: How resilient is the system's end-to-end performance across different underlying LLM backends, and what are the resulting trade-offs between reproduction accuracy and operational cost?
\end{itemize}

\subsection{Datasets}

We employ two distinct datasets to support training and evaluation, summarized in Table~\ref{tab:dataset_overview}. These datasets are designed to assess both controlled benchmarking and real-world applicability.

\begin{table}[t]
\centering
\caption{Overview of Datasets}
\label{tab:dataset_overview}
\resizebox{0.8\columnwidth}{!}{
\begin{tabular}{lccc}
\toprule
\textbf{Dataset} & \textbf{Projects} & \textbf{CWEs} & \textbf{Vulnerabilities} \\
\midrule
FL-Bench-100 & 81 & 4 & 100 \\
GHSA-Real80 & 73 & 9 & 80 \\
\bottomrule
\end{tabular}
}
\end{table}

\subsubsection{FL-Bench-100}
This standardized benchmark consists of 100 confirmed vulnerabilities aggregated from CWE-Bench-Java (real-world Java projects with verified vulnerabilities) \cite{chen2024iris} and PrimeVul (large-scale C/C++ vulnerabilities from public sources) \cite{ding2024vulnerability}. It serves dual purposes: fair comparison against baselines like FaultLine and generating trajectories for adaptive policy training. The distribution across CWE categories is detailed in Table~\ref{tab:cwe_distribution_fl_bench}.

\begin{table}[H]
\centering
\caption{Distribution of Vulnerabilities by CWE in FL-Bench-100}
\label{tab:cwe_distribution_fl_bench}
\resizebox{0.8\columnwidth}{!}{
\begin{tabular}{lccc}
\toprule
\textbf{CWE Group} & \textbf{CWE-Bench-Java} & \textbf{PrimeVul} & \textbf{Total} \\
\midrule
Path Traversal (CWE-22) & 35 & 14 & 49 \\
Command Injection (CWE-78) & 6 & 7 & 13 \\
Cross-Site Scripting (CWE-79) & 15 & 2 & 17 \\
Code Injection (CWE-94) & 14 & 7 & 21 \\
\midrule
\textbf{Total} & \textbf{70} & \textbf{30} & \textbf{100} \\
\bottomrule
\end{tabular}
}
\end{table}

\subsubsection{GHSA-Real80}
We curated this real-world dataset from 80 recent GitHub Security Advisories (GHSA) \cite{githubadvisories}, spanning 73 repositories and 7 programming languages. It emphasizes recency (majority from 2025) to evaluate generalization in practical scenarios. Data collection involved API extraction, stratified sampling by CWE and severity, and manual validation for reproducibility. Statistics by CWE and severity are shown in Table~\ref{tab:db_overview}.

\begin{table}[H]
\centering
\caption{Statistics by CWE and Severity in GHSA-Real80}
\label{tab:db_overview}
\resizebox{\columnwidth}{!}{
\begin{tabular}{lclc}
\toprule
\multicolumn{2}{c}{\textbf{By CWE Group}} & \multicolumn{2}{c}{\textbf{By Severity}} \\
\midrule
\textbf{CWE Group} & \textbf{Count} & \textbf{Severity} & \textbf{Count} \\
\midrule
Path Traversal (CWE-22) & 18 & CRITICAL & 29 \\
Command Injection (CWE-78) & 13 & HIGH & 28 \\
ReDoS (CWE-1333) & 13 & MEDIUM & 23 \\
Cross-Site Scripting (CWE-79) & 9 & & \\
SQL Injection (CWE-89) & 9 & & \\
Deserialization (CWE-502) & 8 & & \\
Input Validation (CWE-20) & 5 & & \\
Prototype Pollution (CWE-1321) & 4 & & \\
SSRF (CWE-918) & 1 & & \\
\midrule
\multicolumn{4}{c}{\textbf{Total Samples: 80}} \\
\bottomrule
\end{tabular}
}
\end{table}

\subsubsection{Data for RL Training}
Trajectories from FL-Bench-100 are split into non-overlapping train/validation (75 CVEs, 86 episodes) and test sets (56 CVEs, 59 episodes) at the CVE level to prevent leakage. CWE distribution for RL data is in Table~\ref{tab:rl_cwe_distribution}.

\begin{table}[H]
\centering
\caption{Distribution of Vulnerabilities by CWE in RL Training Data}
\label{tab:rl_cwe_distribution}
\resizebox{0.8\columnwidth}{!}{
\begin{tabular}{lccc}
\toprule
\textbf{CWE Group} & \textbf{Train+Val} & \textbf{Test} & \textbf{Total} \\
\midrule
Path Traversal (CWE-22) & 50 & 30 & 80 \\
Cross-Site Scripting (CWE-79) & 8 & 10 & 18 \\
Code Injection (CWE-94) & 8 & 9 & 17 \\
Command Injection (CWE-78) & 9 & 7 & 16 \\
\midrule
\textbf{Total} & \textbf{75} & \textbf{56} & \textbf{131} \\
\bottomrule
\end{tabular}
}
\end{table}

\subsection{Implementation}

\begin{figure*}[]
    \centering
    \includegraphics[width=1.8\columnwidth]{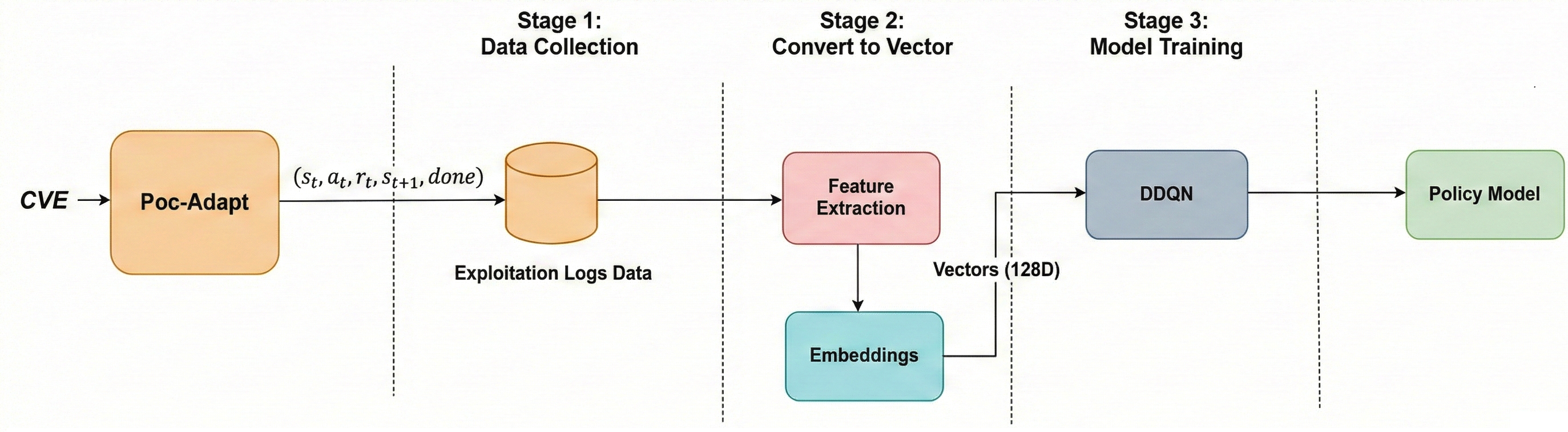} 
    \caption{Policy model training pipeline.}
    \label{fig:rl_training}
\end{figure*}

PoC-Adapt is implemented in Python 3.11.14 using a stateful graph-based architecture orchestrated by LangGraph (v1.0.1) and LangChain (v0.3.27). This design enables persistent memory across iterations and seamless switching between LLM backends (defaulting to Gemini-2.5-Pro with temperature $0.0$ for deterministic outputs). The Adaptive Policy Learning module is built using PyTorch (v2.9.1), trained offline via a Double Deep Q-Network (DDQN) on historical exploitation logs, illustrated in Fig.~\ref{fig:rl_training}. The RL model encodes 13 state features into a 128-dimensional latent vector to approximate Q-values across 14 discrete actions. The model is trained using the Adam optimizer (learning rate $0.001$, batch size $64$, $\gamma=0.99$) and Huber loss ($\delta=1.0$) for 45 epochs, drawing from a replay buffer of 10,000 experiences. During inference, the trained policy dynamically guides the Exploiter agent by recommending top-$k$ actions at each step, strictly bounding the LLM's tool-calling space.

\subsection{Evaluation Metrics}

We employ a set of metrics to comprehensively assess PoC-Adapt's performance at both the system and policy levels. These metrics are selected for their relevance to vulnerability reproduction tasks: they capture success probability, efficiency in resource utilization, and convergence speed, which are critical for evaluating automated exploit generation systems in resource-constrained environments. Below, we describe each metric, its significance, and computation method.

\subsubsection{System-level Metrics}
These metrics evaluate the end-to-end performance of the pipeline, focusing on reproduction success and operational costs.

\begin{itemize}
    \item Success Rate (SR): Measures the proportion of vulnerabilities successfully reproduced. This is the primary indicator of overall effectiveness, as it directly reflects the system's ability to generate verifiable PoCs from vulnerability reports. High SR indicates robust generalization across diverse vulnerabilities.
    \begin{equation}
        \text{SR} = \frac{|\mathcal{S}|}{N} \times 100\%
        \label{eq:sr}
    \end{equation}
    where $\mathcal{S}$ is the set of successfully reproduced vulnerabilities, and $N$ is the total number tested.

    \item Time-to-Exploit (TTE): Computes the average number of action steps required for successful reproductions. TTE quantifies convergence speed, highlighting efficiency in navigating the exploitation process; lower values indicate faster identification of viable PoCs, which is essential for timely vulnerability assessment.
    \begin{equation}
        \text{TTE} = \frac{\sum_{i \in \mathcal{S}} A_i}{|\mathcal{S}|}
        \label{eq:tte}
    \end{equation}
    where $A_i$ is the number of steps for the $i$-th successful reproduction.

    \item Exploit Efficiency (EE): Represents the number of successes per total action across all attempts. EE assesses resource optimization, penalizing inefficient trials in failures; higher EE signifies better action utilization, crucial for scaling to large vulnerability sets.
    \begin{equation}
        \text{EE} = \frac{|\mathcal{S}|}{\sum_{i=1}^{N} A_i}
        \label{eq:ee}
    \end{equation}

    \item Token Consumption (k tokens): Total LLM tokens (input + output) per vulnerability attempt, reported in thousands (k). This metric evaluates computational overhead, as token usage correlates with inference costs and latency; it is vital for assessing practicality in budget-limited deployments.

    \item Monetary Cost (\$): Inference cost in USD, derived from token consumption using public LLM pricing. This provides a real-world economic perspective, enabling trade-off analysis between performance and operational expenses.
\end{itemize}

\subsubsection{Policy-level Metrics}
These metrics specifically evaluate the adaptive policy learning component, focusing on decision-making quality in the exploitation phase.

\begin{itemize}
    \item Policy Success Rate (SR$_{\text{policy}}$): Proportion of episodes (trajectories) ending in successful reproduction. This measures the policy's ability to guide the agent to verifiable PoCs, indicating learned exploitation strategies' effectiveness over random actions.
    \begin{equation}
        \text{SR}_{\text{policy}} = \frac{|\mathcal{E}|}{M} \times 100\%
    \end{equation}
    where $\mathcal{E}$ is the set of successful episodes, and $M$ is the total episodes.

    \item Policy Time-to-Exploit (TTE$_{\text{policy}}$): Average steps to success across successful episodes. TTE$_{\text{policy}}$ assesses convergence efficiency under the policy, with lower values showing the policy's skill in selecting optimal actions based on observed states.
    \begin{equation}
        \text{TTE}_{\text{policy}} = \frac{\sum_{i \in \mathcal{E}} A_i}{|\mathcal{E}|}
    \end{equation}

    \item Policy Exploit Efficiency (EE$_{\text{policy}}$): Successes per total action across all episodes. EE$_{\text{policy}}$ evaluates action economy, rewarding policies that minimize wasteful steps; it is key for validating the policy's adaptability in large action spaces.
    \begin{equation}
        \text{EE}_{\text{policy}} = \frac{|\mathcal{E}|}{\sum_{i=1}^{M} A_i}
    \end{equation}
\end{itemize}


\section{Results and Analysis}
\label{sec:results}

\subsection{Answer to RQ1: Effectiveness on FL-Bench-100}
\label{sec:rq1}

To evaluate RQ1, we compare PoC-Adapt against the state-of-the-art baseline FaultLine on the standardized FL-Bench-100 benchmark under identical constraints: a \$1 inference budget, 60-minute timeout, and maximum 3 refinement loops per vulnerability. Also, the external tools for LLM agents are setted up with the same configurations for tool calling. This setup ensures a fair assessment of reproduction effectiveness.

Table~\ref{tab:sr_compare} summarizes the results. PoC-Adapt achieves a success rate (SR) of 15\% (15/100 vulnerabilities), outperforming FaultLine's 12\% (12/100), representing a 25\% relative improvement. This gain highlights PoC-Adapt's enhanced ability to generate verifiable PoCs within limited resources.

Moreover, PoC-Adapt demonstrates superior efficiency: Time-to-Exploit (TTE) is halved (16.33 vs. 35.92 steps), indicating faster convergence to successful reproductions. Exploit Efficiency (EE) is more than doubled (0.025 vs. 0.011), reflecting better resource utilization across attempts. These improvements stem from PoC-Adapt's multi-agent coordination, semantic verification, and adaptive policy, which reduce heuristic trial-and-error compared to FaultLine's hierarchical reasoning.

\begin{table}[H]
\centering
\caption{Comparison with FaultLine on FL-Bench-100}
\label{tab:sr_compare}
\resizebox{0.8\columnwidth}{!}{
\begin{tabular}{lccc}
\toprule
\textbf{System} & \textbf{SR (\%)} & \textbf{TTE (steps)} & \textbf{EE} \\
\midrule
FaultLine & 12.0 & 35.92 & 0.011 \\
PoC-Adapt (Ours) & \textbf{15.0} & \textbf{16.33} & \textbf{0.025} \\
\bottomrule
\end{tabular}
}
\end{table}

\begin{tcolorbox}[title={Answer to RQ1}]
PoC-Adapt outperforms FaultLine on FL-Bench-100 with a 25\% relative SR improvement (15\% vs. 12\%), halved TTE, and doubled EE, demonstrating superior effectiveness and efficiency in vulnerability reproduction under identical constraints.
\end{tcolorbox}

\subsection{Answer to RQ2: Practicality on GHSA-Real80}
\label{sec:rq2}

For RQ2, we assess PoC-Adapt's end-to-end reproduction performance on the real-world GHSA-Real80 dataset and identify failure bottlenecks across pipeline stages.

PoC-Adapt successfully reproduces 12 out of 80 vulnerabilities, yielding an SR of 15\%. This rate underscores the challenges of real-world advisories, which often lack detailed exploitation contexts, yet affirms PoC-Adapt's practical viability.

\begin{figure}[H]
    \centering
    \includegraphics[width=\columnwidth]{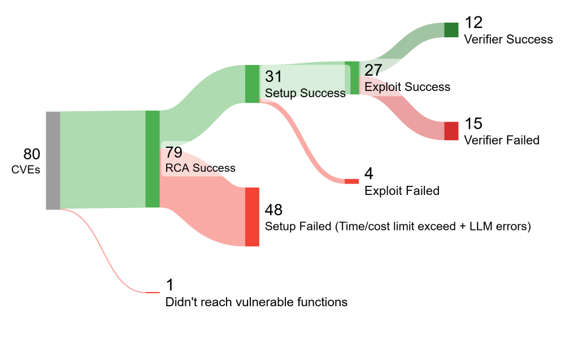} 
    \caption{Detailed stage-wise analysis of PoC-Adapt on GHSA-Real80.}
    \label{fig:failure_analyze}
\end{figure}

Failure analysis, detailed in 
Fig.~\ref{fig:failure_analyze}, reveals the Planner stage as the primary bottleneck with a 60.76\% conditional failure rate (48/79 cases). This highlights difficulties in environment setup due to diverse dependencies and configurations. The Exploiter stage is more robust (12.90\% failure), while Validator filters out 55.56\% of candidates, reducing false positives but indicating room for refined semantic checks.


\begin{tcolorbox}[title={Answer to RQ2}]
PoC-Adapt achieves 15\% SR on GHSA-Real80, successfully reproducing 12 vulnerabilities. Failures concentrate in Planner (60.76\%), underscoring environment setup as the main challenge in real-world deployment.
\end{tcolorbox}

\subsection{Answer to RQ3: Generalizability across CWE and Severity Levels}
\label{sec:rq3}

RQ3 examines reproduction variance by CWE categories and severity levels on GHSA-Real80 to assess generalization.

\begin{figure}[H]
    \centering
    \includegraphics[width=\columnwidth]{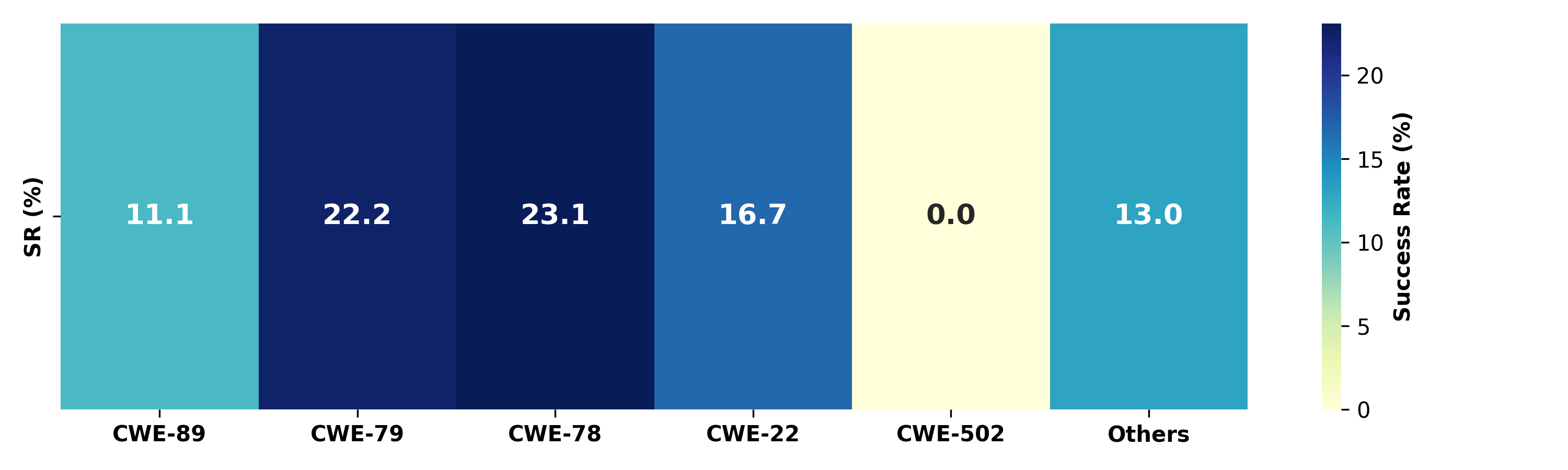} 
    \caption{Success Rate (SR) by CWE on GHSA-Real80}
    \label{fig:cwe_success_rate}
\end{figure}

Fig.~\ref{fig:cwe_success_rate} shows highest SR for CWE-78 (Command Injection, 23.1\%) and CWE-79 (XSS, 22.2\%), with CWE-22 (Path Traversal) at 16.7\%. CWE-502 (Deserialization) yields 0\%, indicating limitations in handling context-dependent vulnerabilities. This suggests PoC-Adapt excels on direct-impact web vulnerabilities but struggles with subtle, logic-based ones.


By severity (Table~\ref{tab:severity_distribution}), Critical and High levels achieve 20.7\% and 17.9\% SR, respectively, vs. 4.3\% for Medium. High-impact vulnerabilities provide clearer runtime signals, aligning with the semantic oracle's strengths, while Medium ones often require nuanced preconditions.

\begin{table}[H]
\centering
\caption{SR by Severity on GHSA-Real80}
\label{tab:severity_distribution}
\resizebox{0.6\columnwidth}{!}{
\begin{tabular}{lccc}
\toprule
\textbf{Severity} & Critical & High & Medium \\
\midrule
\textbf{SR (\%)} & \textbf{20.7} & 17.9 & 4.3 \\
\bottomrule
\end{tabular}
}
\end{table}

\begin{tcolorbox}[title={Answer to RQ3}]
PoC-Adapt generalizes across 6/9 CWE categories with up to 23.1\% SR, performing best on direct-impact vulnerabilities. Higher severity levels yield better results (20.7\% for Critical), suggesting improvements needed for subtle, medium-severity cases.
\end{tcolorbox}

\subsection{Answer to RQ4: Efficiency and Cost Analysis}
\label{sec:rq4}

RQ4 analyzes computational costs on GHSA-Real80, including time, tokens, and monetary equivalents, with breakdown by stage.

Successful reproductions average 7.14 minutes (\$0.42, 320.2k tokens), while failures take 13.87 minutes (\$0.64, 499.2k tokens). Failures consume more due to prolonged refinement loops. Planner and Exploiter dominate token usage, as they involve extensive dependency resolution and iterative PoC generation.

\begin{tcolorbox}[title={Answer to RQ4}]
PoC-Adapt incurs an average cost of \$0.42 and 7.14 minutes per successful reproduction, with Planner and Exploiter as primary token consumers, highlighting setup and exploitation as efficiency bottlenecks.
\end{tcolorbox}

\subsection{Answer to RQ5: Contribution of Adaptive Policy Learning}
\label{sec:rq5}

For RQ5, we evaluate the policy on a held-out test set (59 episodes, max 50 steps/episode) against a random agent baseline (averaged over 100 seeds).

Table~\ref{tab:rl_policy_metrics} shows DDQN achieves 44.83\% SR$_{\text{policy}}$ (vs. 38.86\%), halves TTE (22.19 vs. 40.73 steps), and quintuples EE (0.0427 vs. 0.0084), confirming learned state-action relationships over randomness.
\begin{table}[H]
\centering
\caption{DDQN vs. Random Agent on FL-Bench-100 Test Set}
\label{tab:rl_policy_metrics}
\resizebox{0.8\columnwidth}{!}{
\begin{tabular}{lccc}
\toprule
\textbf{Algorithm} & \textbf{SR$_{\text{policy}}$ (\%)} & \textbf{TTE$_{\text{policy}}$ (steps)} & \textbf{EE$_{\text{policy}}$} \\
\midrule
Random Agent & 38.86 $\pm$ 4.18 & 40.73 $\pm$ 0.99 & 0.0084 $\pm$ 0.0010 \\
DDQN (Ours) & \textbf{44.83} & \textbf{22.19} & \textbf{0.0427} \\
\bottomrule
\end{tabular}
}
\end{table}

Ablation on GHSA-Real80 (Table~\ref{tab:pipeline_rl_comparison}) reveals RL integration boosts SR to 17.5\% (+16.7\% relative) and improves EE, albeit with 23\% higher tokens due to policy inference overhead.

\begin{table}[H]
\centering
\caption{Ablation Study on GHSA-Real80}
\label{tab:pipeline_rl_comparison}
\resizebox{0.6\columnwidth}{!}{
\begin{tabular}{lcccc}
\toprule
\textbf{Metric} & \textbf{Without RL} & \textbf{With RL} \\
\midrule
SR (\%) & 15.0 & \textbf{17.5} \\
TTE (steps) & 20.67 & \textbf{19.43} \\
Tokens (k) & \textbf{320.2} & 393.4 \\
EE & 0.048 & \textbf{0.052} \\
\bottomrule
\end{tabular}
}
\end{table}

\begin{tcolorbox}[title={Answer to RQ5}]
The adaptive policy (DDQN) significantly outperforms random baselines, halving TTE and improving SR by 15\%. Ablation confirms a 16.7\% relative SR gain with modest cost increase, validating its contribution to adaptability.
\end{tcolorbox}

\subsection{Answer to RQ6: Robustness across LLM Backends}
\label{sec:rq6}

RQ6 investigates performance variance when replacing Gemini-2.5-Pro with alternative backends (Table~\ref{tab:llm_specs}), keeping all other components fixed.

On FL-Bench-100 and GHSA-Real80 (Table~\ref{tab:llm_backend_compare}), Gemini yields the highest SR (15\%) but at elevated costs (Table~\ref{tab:llm_cost}). DeepSeek-V3 minimizes cost (\$0.07) and time (3.59 min) but reduces SR to 6.25\%. Qwen-3 balances with 11.25\% SR at \$0.13, though with longer processing (15.65 min).

\begin{table}[H]
\centering
\caption{LLM Specifications}
\label{tab:llm_specs}
\resizebox{\columnwidth}{!}{
\begin{tabular}{lccccc}
\toprule
\textbf{LLM} & \textbf{Provider} & \textbf{Params} & \textbf{Context} & \textbf{\$ /1M Input} & \textbf{\$ /1M Output} \\
\midrule
Gemini-2.5-Pro & Google Cloud & -- & 1M & 1.25 & 10.00 \\
DeepSeek-V3 & DeepInfra & 671B (A37B) & 160K & 0.27 & 0.89 \\
Qwen-3 & DeepInfra & 235B (A22B) & 256K & 0.23 & 2.39 \\
\bottomrule
\end{tabular}
}
\end{table}

These trade-offs indicate PoC-Adapt's robustness, with performance scaling to model capability while maintaining pipeline integrity.

\begin{table}[H]
\centering
\caption{Performance across LLM Backends}
\label{tab:llm_backend_compare}
\resizebox{\columnwidth}{!}{
\begin{tabular}{lccccc}
\toprule
\textbf{Dataset} & \textbf{LLM} & \textbf{SR (\%)} & \textbf{TTE (steps)} & \textbf{EE} \\
\midrule
\multirow{3}{*}{FL-Bench-100} & Gemini-2.5-Pro & \textbf{15.0} & \textbf{16.33} & 0.025 \\
& DeepSeek-V3 & 10.0 & 22.30 & 0.024 \\
& Qwen-3 & 12.0 & 17.50 & \textbf{0.028} \\
\midrule
\multirow{3}{*}{GHSA-Real80} & Gemini-2.5-Pro & \textbf{15.0} & 20.67 & \textbf{0.048} \\
& DeepSeek-V3 & 6.25 & 26.00 & 0.011 \\
& Qwen-3 & 11.25 & \textbf{16.11} & 0.025 \\
\bottomrule
\end{tabular}
}
\end{table}

\begin{table}[H]
\centering
\caption{Inference Costs across LLM Backends on GHSA-Real80}
\label{tab:llm_cost}
\resizebox{0.8\columnwidth}{!}{
\begin{tabular}{lccc}
\toprule
\textbf{LLM} & \textbf{Avg. Tokens (k)} & \textbf{Avg. Cost (\$)} & \textbf{Avg. Time (min)} \\
\midrule
Gemini-2.5-Pro & 472.4 & 0.61 & 12.86 \\
DeepSeek-V3 & \textbf{249.1} & \textbf{0.07} & \textbf{3.59} \\
Qwen-3 & 279.5 & 0.13 & 15.65 \\
\bottomrule
\end{tabular}
}
\end{table}

\begin{tcolorbox}[title={Answer to RQ6}]
PoC-Adapt maintains consistent performance across LLM backends, with Gemini-2.5-Pro offering the best SR at higher cost, DeepSeek-V3 minimizing expenses, and Qwen-3 providing a balanced trade-off—demonstrating robustness and flexibility.
\end{tcolorbox}

\section{Discussion}
\label{sec:discussion}

This section reflects on the experimental findings, discusses factors influencing performance, highlights limitations, and addresses threats to validity. The results demonstrate that PoC-Adapt advances automated PoC synthesis and verification through multi-agent coordination, semantic state-differencing verification, and adaptive policy learning. Nevertheless, several challenges persist, and important validity concerns must be considered.

\subsection{Limitations}
\label{sec:limitations}

Although PoC-Adapt demonstrates notable improvements over prior LLM-based approaches, several limitations remain and highlight important directions for future work.

The most significant bottleneck lies in environment reproduction. On the GHSA-Real80 dataset, the Planner stage suffers from a high conditional failure rate of 60.76\% (48 out of 79 cases). This is primarily caused by the heterogeneity of real-world repositories, including diverse build systems, dependency conflicts, incomplete or outdated documentation, and version-specific configuration requirements. Even with Docker-based isolation, complex multi-service applications or non-standard build processes frequently exceed the imposed 60-minute timeout or \$1 budget, resulting in premature termination of the pipeline.

Another key limitation concerns the sensitivity of the Semantic Oracle. While the oracle performs well on high-signal, direct-impact vulnerabilities such as command injection (23.1\% SR) and cross-site scripting (22.2\% SR), its effectiveness drops considerably for subtle or context-dependent cases. In particular, success rate falls to 0\% for deserialization vulnerabilities (CWE-502) and only 4.3\% for medium-severity vulnerabilities. Ambiguous state changes or indirect impacts are difficult to capture reliably within the strict refinement budget ($B=3$), leading to overly conservative filtering and potential false negatives.

Computational cost and heavy reliance on powerful LLMs also present practical challenges. Successful reproductions with Gemini-2.5-Pro require an average of \$0.42 and 320.2k tokens, while failed attempts incur higher costs due to prolonged refinement loops. Substitution experiments reveal clear trade-offs: although cheaper models such as DeepSeek-V3 significantly reduce cost (to \$0.07), they degrade the overall success rate to 6.25\%, highlighting the system's current dependence on high-capability LLMs for reasoning-intensive stages including root cause analysis, environment planning, and exploit generation.

Furthermore, the adaptive policy learning mechanism is trained on a relatively small set of only 86 trajectories from FL-Bench-100. Although ablation studies confirm a 16.7\% relative improvement in success rate, the limited number of positive samples in certain CWE categories restricts the policy's ability to generalize effectively to the more diverse scenarios in GHSA-Real80.

Finally, the fixed refinement budget ($B=3$) represents a deliberate trade-off between cost and quality. However, this constraint may prematurely terminate promising exploits that require additional iterations, especially in complex or poorly documented environments.

\subsection{Threats to Validity}
\label{sec:threats_to_validity}

We categorize threats to validity into internal and external concerns that could influence the interpretation and generalizability of our results.

\subsubsection{Internal Validity}

\paragraph{LLM non-determinism and experimental reproducibility}
Internal validity concerns whether the observed performance improvements can be confidently attributed to the proposed mechanisms rather than confounding factors. Despite fixing the temperature at 0.0, inherent LLM stochasticity and potential API-level non-determinism still introduce variability. Although consistent random seeds and multiple independent runs were employed to stabilize results, residual randomness may affect exact reproducibility in certain cases.

\paragraph{RL policy training and reward design}
The adaptive policy was trained exclusively on trajectories from FL-Bench-100 using CVE-disjoint train/test splits to prevent direct data leakage. While shared CWE patterns across splits could theoretically introduce subtle indirect leakage, we consider this risk negligible due to the high diversity of exploitation strategies in real-world scenarios. Additionally, the reward function combines dense step-wise and terminal rewards through heuristic design. Ablation studies confirm its positive contribution; however, we have not exhaustively explored all possible alternative reward-shaping strategies.

\paragraph{Semantic Oracle and state representation}
The Semantic Oracle depends on predefined system state observations (e.g., file hashes, environment variables, and database records) extracted from the impact hypothesis. A potential threat is that certain edge-case exploits may produce delayed side effects or out-of-band behaviors (such as network exfiltration) that our sandbox does not fully profile during state differencing, possibly leading to false negatives. Furthermore, extracting the 13 state features from raw execution logs for the RL agent relies on heuristic parsing. Although this process was standardized, subtle contextual nuances in the logs may be lost, potentially affecting the quality of policy learning.

\paragraph{Operational constraints and construct validity}
We imposed strict operational constraints (\$1 budget, 60-minute timeout, and at most $B=3$ refinement loops) to simulate realistic deployment conditions. While practical, these bounds may disadvantage slower yet ultimately correct exploitation strategies, thus conditioning the results on resource-limited scenarios. Regarding construct validity, although Time-to-Exploit (TTE) and Exploit Efficiency (EE) effectively reflect operational cost and efficiency, they may not fully capture the intrinsic difficulty or complexity of the vulnerabilities themselves. An exploit with high TTE might simply result from complex environment setup rather than sophisticated exploitation logic.

These internal validity threats are carefully considered in our experimental design and are mitigated where possible through ablation studies, multiple runs, and controlled constraints.

\subsubsection{External Validity}

\paragraph{Generalizability across vulnerability datasets}
External validity concerns the extent to which our findings generalize beyond the experimental conditions. Although FL-Bench-100 was carefully curated for benchmarking, it may not fully capture the heterogeneity of in-the-wild vulnerabilities. GHSA-Real80 improves realism by using disclosed, patch-available GitHub advisories; however, it still excludes zero-day vulnerabilities, proprietary codebases, and cases with ambiguous or missing reproduction steps.

\paragraph{Dependency on LLM backends and infrastructure}
The observed performance is tightly coupled with current-generation LLM backends (primarily Gemini-2.5-Pro, with additional tests on DeepSeek-V3 and Qwen-3). Future models with longer context windows, stronger reasoning capabilities, or lower inference costs could significantly alter the reported trade-offs. Similarly, all experiments were conducted on a single workstation using Docker containers; results may differ under distributed cloud environments, alternative operating systems, container runtimes, or more restrictive hardened sandboxes.

\paragraph{Limited coverage of vulnerability types}
Although GHSA-Real80 spans nine CWE categories and multiple severity levels, certain complex classes such as memory corruption and race conditions remain underrepresented. This limits broad claims regarding the framework’s applicability across all vulnerability types.

\paragraph{Baselines and experimental constraints}
Our controlled benchmarking focused primarily on FaultLine due to the lack of publicly available, reproducible implementations of other recent multi-agent frameworks (e.g., CVE-Genie or PTFusion) that align with the FL-Bench-100 setup. Furthermore, experiments enforced a strict 60-minute timeout, \$1 budget limit, and maximum of three refinement loops. While these constraints mirror practical operational budgets in CI/CD pipelines, real-world adversaries or expert analysts often operate with substantially larger time and resource allowances. Consequently, the reported success rates reflect efficiency under constrained conditions rather than the theoretical upper bound of the system. Finally, reliance on commercial LLM APIs introduces the risk of ``API drift,'' whereby silent model updates by providers may subtly affect behavior and reproducibility over time, even with temperature set to 0.0.

These external validity threats are inherent to contemporary LLM-driven automated exploit generation research and directly motivate the future research directions outlined in Section~\ref{sec:future_directions}.

\subsection{Ethical Considerations}

This work exclusively targets publicly disclosed and already-patched vulnerabilities from standardized benchmarks and real-world GHSA advisories. All experiments were performed in isolated Docker sandboxes with strict least-privilege tool allocation and automatic environment destruction after each trial. No production systems or unpatched vulnerabilities were accessed. The primary goal of PoC-Adapt is defensive: to reduce the reproducibility gap in vulnerability management by enabling more reliable root-cause analysis, environment setup, and semantic verification of exploit impact. By improving verification accuracy and lowering generation cost, the framework assists security teams and open-source maintainers in faster risk assessment and patch validation.

We acknowledge the dual-use potential of LLM-based multi-agent systems combined with adaptive policy learning for automated exploit generation. Such techniques could lower the barrier for malicious actors to weaponize vulnerabilities. To mitigate these risks, we enforced controlled tool access, inter-agent feedback loops, bounded refinement iterations, and a semantic state-differencing oracle that requires strict matching between hypothesized and observed system states. All generated PoCs are intended solely for research and defensive purposes. We encourage the community to apply PoC-Adapt only within coordinated vulnerability disclosure frameworks and to prioritize defensive applications such as automated regression testing and patch quality assessment. This research adheres to the ACM Code of Ethics and IEEE principles on responsible conduct in offensive security and AI.

\subsection{Future Directions}
\label{sec:future_directions}

Several promising directions emerge from the limitations identified in this study. The predominant failure in environment reproduction (Planner stage) motivates research into task decomposition techniques that break setup into verifiable sub-tasks with automated dependency resolution and fallback strategies, potentially integrating reinforcement learning for adaptive configuration planning. To address challenges with subtle, context-dependent vulnerabilities, incorporating Retrieval-Augmented Generation (RAG) would enrich the agents' domain knowledge by retrieving relevant external analyses, exploit patterns, or framework-specific documentation, thereby reducing ambiguity in root cause analysis and hypothesis formulation. Extending coverage to UI-driven exploits requires integrating browser automation tools such as Playwright or Selenium, enabling agents to simulate user interactions and trigger vulnerabilities in web applications that cannot be reproduced via API or command-line alone. For long-term adaptability, transitioning from purely offline policy learning to periodic fine-tuning or online updates using newly collected trajectories would allow the DDQN policy to continuously evolve in response to emerging vulnerability patterns and changing exploitation environments. Finally, deploying on-premise LLMs with targeted fine-tuning on security-specific corpora would mitigate API dependency, reduce latency and cost variability, enhance operational stability, and improve data privacy in enterprise or sensitive settings. Pursuing these advancements will strengthen PoC-Adapt's robustness, scalability, and practical utility for real-world vulnerability management workflows.
\section{Conclusion}
\label{sec:conclusion}

The lack of verifiable Proof-of-Concept (PoC) exploits creates a critical reproducibility gap in vulnerability management, impeding accurate exploitability assessment and timely remediation. Existing LLM-based approaches often rely on superficial oracles and heuristic trial-and-error, resulting in unreliable outputs and high computational inefficiency in complex scenarios. 

This work introduces PoC-Adapt, an end-to-end multi-agent framework that fundamentally shifts the automated exploit generation (AEG) paradigm. By integrating semantic state-differencing verification with adaptive policy learning via offline DDQN training, PoC-Adapt transitions LLM agents from blind heuristic exploration to state-aware, optimized reasoning. Experimental results on FL-Bench-100 show a 25\% relative success rate improvement (15\% vs. 12\% for FaultLine), halved time-to-exploit (16.33 vs. 35.92 steps), and doubled exploit efficiency, while on the real-world GHSA-Real80 dataset, PoC-Adapt reproduces 12 vulnerabilities across 6/9 CWE categories at 15\% success rate and \$0.42 average cost of merely per success. The adaptive policy significantly outperforms random baselines (44.83\% vs. 38.86\% policy success rate), with ablation confirming a 16.7\% relative gain, and performance remains robust across LLM backends. These findings confirm that modeling the exploitation process as a Markov Decision Process (MDP) effectively eliminates the ``token explosion" problem, making LLM-driven AEG economically viable for large-scale security operations. Furthermore, the Semantic Oracle strictly guarantees the validity of generated exploits, successfully neutralizing LLM hallucinations.

Despite these methodological advances, our detailed failure analysis identifies environment setup as the dominant bottleneck in real-world deployments. Consequently, future work should focus on improving environment reproduction through task decomposition, incorporating Retrieval-Augmented Generation for richer domain knowledge, extending to UI-driven exploits, enabling online policy updates, and transitioning to on-premise LLMs to reduce dependency, latency, and privacy risks, thereby advancing PoC-Adapt toward scalable deployment in real-world vulnerability workflows. Ultimately, PoC-Adapt paves the way for a transition from reactive vulnerability scanning to continuous, automated, and verifiable risk validation.

\section*{Acknowledgement}

This research is funded by Vietnam National University Ho Chi Minh City (VNU-HCM) under grant number NCM2025-26-01.









\bibliographystyle{unsrt}

\bibliography{refs.bib}

\end{document}